\documentclass{PoS}

\title{Search for the dark photon in $\pi^0$ decays}

\ShortTitle{Search for the dark photon in $\pi^0$ decays}

\author{\speaker{Evgueni Goudzovski}%
         \thanks{For the NA48/2 Collaboration: G.~Anzivino, R.~Arcidiacono, W.~Baldini, S.~Balev, J.R.~Batley, M.~Behler, S.~Bifani, C.~Biino, A.~Bizzeti, B.~Bloch-Devaux
% ...
}\\
        School of Physics and Astronomy, University of Birmingham, B15 2TT, United Kingdom\\
        E-mail: \email{eg@hep.ph.bham.ac.uk.}}

\abstract{A sample of $1.69\times 10^7$ fully reconstructed $\pi^0\to\gamma e^+e^-$ decay candidates collected by the NA48/2 experiment at CERN in 2003--2004 is analysed to search for the dark photon ($A'$) production in the $\pi^0\to\gamma A'$ decay followed by the prompt $A'\to e^+e^-$ decay. No signal is observed, and an exclusion region in the plane of the dark photon mass $m_{A'}$ and mixing parameter $\varepsilon^2$ is established. The obtained upper limits on $\varepsilon^2$ are more stringent than the previous limits in the mass range $9~{\rm MeV}/c^2<m_{A'}<70~{\rm MeV}/c^2$. The prospects of the dark photon search in the $K^\pm\to\pi^\pm A'$ decay are also discussed.}

\FullConference{18th International Conference From the Planck Scale to the Electroweak Scale \\
		 25-29 May 2015\\
		 Ioannina, Greece }

\begin{document}

\section*{Introduction}

Kaons are a source of tagged neutral pion decays, and high intensity kaon experiments provide opportunities for precision $\pi^0$ decay measurements. The NA48/2 experiment at the CERN SPS collected a large sample of charged kaon ($K^\pm$) decays in flight, corresponding to about $2\times 10^{11}$ $K^\pm$ decays in the fiducial decay volume. The search for a hypothetical dark photon (DP, denoted $A'$) using a large sample of tagged $\pi^0$ mesons from $K^\pm\to\pi^\pm\pi^0$ and $K^\pm\to\pi^0\mu^\pm\nu$ decays~\cite{ba15} is presented, and the prospects for dark photon search in the $K^\pm\to\pi^\pm A'$ decay are discussed.

In a rather general set of hidden sector models with an extra $U(1)$ gauge symmetry~\cite{ho86}, the interaction of the DP with the visible sector proceeds through kinetic mixing with the Standard Model (SM) hypercharge. Such scenarios with GeV-scale dark matter provide possible explanations to the observed rise in the cosmic-ray positron fraction with energy and the muon gyromagnetic ratio $(g-2)$ measurement~\cite{po09}. The DP is characterized by two a priori unknown parameters, the mass $m_{A'}$ and the mixing parameter $\varepsilon^2$. Its possible production in the $\pi^0$ decay and its subsequent decay proceed via the chain $\pi^0\to\gamma A'$, $A'\to e^+e^-$. The expected branching fraction of the above $\pi^0$ decay is~\cite{batell09}
\begin{equation}
{\cal B}(\pi^0\to\gamma A') = 2\varepsilon^2 \left(1-\frac{m_{A'}^2}{m_{\pi^0}^2}\right)^3 {\cal B}(\pi^0\to\gamma\gamma),
\label{eq:br}
\end{equation}
which is kinematically suppressed as $m_{A'}$ approaches $m_{\pi^0}$. In the DP mass range $2m_e<m_{A'}<m_{\pi^0}$ accessible in pion decays, the only allowed tree-level decay into SM fermions is $A'\to e^+e^-$, while the loop-induced SM decays ($A'\to 3\gamma$, $A'\to\nu\bar\nu$) are highly suppressed. Therefore, for a DP decaying only into SM particles, ${\cal B}(A'\to e^+e^-)\approx 1$, and the expected total decay width is~\cite{batell09}
\begin{equation}
\Gamma_{A'} \approx \Gamma(A'\to e^+e^-) = \frac{1}{3} \alpha\varepsilon^2 m_{A'} \sqrt{1-\frac{4m_e^2}{m_{A'}^2}}\left(1+\frac{2m_e^2}{m_{A'}^2}\right).
\end{equation}
It follows that, for $2m_e\ll m_{A'}<m_{\pi^0}$, the DP mean proper lifetime $\tau_{A'}$ satisfies the relation
\begin{equation}
c\tau_{A'} = \hbar c / \Gamma_{A'} \approx 0.8~{\mu\rm m} \times \left(\frac{10^{-6}}{\varepsilon^2}\right) \times \left(\frac{100~{\rm MeV}/c^2}{m_{A'}}\right).
\end{equation}
This analysis is performed assuming that the DP decays at the production point ({\it prompt decay}), which is valid for sufficiently large values of $m_{A'}$ and $\varepsilon^2$. In this case, the DP production and decay signature is identical to that of the Dalitz decay $\pi^0_D\to e^+e^-\gamma$, which therefore represents an irreducible background and determines the sensitivity. Pure $\pi^0_D$ decay samples are reconstructed through the $K^\pm\to\pi^\pm\pi^0$ and $K^\pm\to\pi^0\mu^\pm\nu$ decays (denoted $K_{2\pi}$ and $K_{\mu 3}$). Additionally, the $K^\pm\to\pi^\pm\pi^0\pi^0$ decay (denoted $K_{3\pi}$) is considered as a background in the $K_{\mu 3}$ sample.

\section{NA48/2 beam, detector and data sample}
\label{sec:experiment}

The NA48/2 experiment used simultaneous $K^+$ and $K^-$ beams produced by 400~GeV/$c$ primary CERN SPS protons impinging on a beryllium target. Charged particles with momenta of $(60\pm3)$ GeV/$c$ were selected by an achromatic system of four dipole magnets which split the two beams in the vertical plane and recombined them on a common axis. The beams then passed through collimators and a series of quadrupole magnets, and entered a 114~m long cylindrical vacuum tank with a diameter of 1.92 to 2.4~m containing the fiducial decay region.

The vacuum tank was followed by a magnetic spectrometer housed in a vessel filled with helium at nearly atmospheric pressure, separated from the vacuum by a thin ($0.3\%~X_0$) $\rm{Kevlar}\textsuperscript{\textregistered}$ window.
An aluminium beam pipe of 158~mm outer diameter traversing the centre of the spectrometer (and all the following detectors) allowed the undecayed beam particles to continue their path in vacuum. The spectrometer consisted of four drift chambers (DCH) with an octagonal transverse width of 2.9~m: DCH1, DCH2 located upstream and DCH3, DCH4 downstream of a dipole magnet that provided a horizontal transverse momentum kick of 120~MeV/$c$ for charged particles. Each DCH was composed of eight planes of sense wires. The DCH space point resolution was 90~$\mu$m in both horizontal and vertical directions, and the momentum resolution was $\sigma_p/p = (1.02 \oplus 0.044\cdot p)\%$, with $p$ expressed in GeV/$c$. The spectrometer was followed by a plastic scintillator hodoscope (HOD) with a transverse size of about 2.4 m, consisting of a plane of vertical and a plane of horizontal strip-shaped counters arranged in four quadrants (each logically divided into four regions). The HOD provided time measurements of charged particles with 150~ps resolution. It was followed by a liquid krypton electromagnetic calorimeter (LKr), an almost homogeneous ionization chamber with an active volume of 7 m$^3$ of liquid krypton, $27~X_0$ deep, segmented transversally into 13248 projective $\sim\!2\!\times\!2$~cm$^2$ cells. The LKr energy resolution was $\sigma_E/E=(3.2/\sqrt{E}\oplus9/E\oplus0.42)\%$, the spatial resolution for an isolated electromagnetic shower was $(4.2/\sqrt{E}\oplus0.6)$~mm in both horizontal and vertical directions, and the time resolution was $2.5~{\rm ns}/\sqrt{E}$, with $E$ expressed in GeV. A detailed description of the beamline and detector is given in Ref.~\cite{fa07,ba07}.

The NA48/2 experiment collected data in 2003--2004, during about 100 days of efficient data taking in total. A two-level trigger chain was employed to collect $K^\pm$ decays with at least three charged tracks in the final state~\cite{ba07}. At the first level (L1), a coincidence of hits in the two planes of the HOD was required to occur in at least two of 16 non-overlapping regions. The second level (L2) performed online reconstruction of trajectories and momenta of charged particles based on the DCH information. The L2 logic was based on the multiplicities and kinematics of reconstructed tracks and two-track vertices.

\boldmath
\section{Simulation of the $\pi^0_D$ background}
\unboldmath
\label{sec:background}

Simulations of the $K_{2\pi}$, $K_{\mu 3}$ and $K_{3\pi}$ decays followed by the $\pi^0_D$ decay (denoted $K_{2\pi D}$, $K_{\mu3 D}$ and $K_{3\pi D}$) are performed to evaluate the integrated kaon flux and to estimate the irreducible $\pi^0_D$ background to the DP signal. The $K_{2\pi}$ and $K_{\mu 3}$ decays are simulated including final-state radiation~\cite{ga06}. The $\pi^0_D$ decay is simulated using the lowest-order differential decay rate~\cite{mi72}
\begin{equation}
\label{eq:dgdxdy}
\frac{d^2\Gamma}{dxdy} = \Gamma_0\frac{\alpha}{\pi}|F(x)|^2\frac{(1-x)^3}{4x}\left(1+y^2+\frac{r^2}{x}\right),
\end{equation}
where $\Gamma_0$ is the $\pi^0\to\gamma\gamma$ decay rate, $r = 2m_e/m_{\pi^0}$, and $F(x)$ is the pion transition form factor (TFF). The kinematic variables are
\begin{equation}
x = \frac{(Q_1+Q_2)^2}{m_{\pi^0}^2} = (m_{ee}/m_{\pi^0})^2, ~~~~~ y = \frac{2P(Q_1-Q_2)}{m_{\pi^0}^2(1-x)},
\end{equation}
where $Q_1$, $Q_2$ and $P$ are the four-momenta of the two electrons ($e^\pm$) and the pion ($\pi^0$), and $m_{ee}$ is the invariant mass of the $e^+e^-$ pair. Radiative corrections to the $\pi^0_D$ decay are implemented following the approach of Mikaelian and Smith~\cite{mi72}: the differential decay rate is modified by a radiative correction factor that depends on the kinematic variables $x$ and $y$.

The TFF is conventionally parameterized as $F(x)=1+ax$. Vector meson dominance models expect the slope parameter to be $a\approx (m_{\pi^0}/m_\rho)^2 \approx 0.03$~\cite{la85}, while detailed calculations based on dispersion theory obtain $a = 0.0307\pm0.0006$~\cite{ho14}. Experimentally, the PDG average value $a=0.032\pm0.004$~\cite{pdg} is dominated by an $e^+e^-\to e^+e^-\pi^0$ measurement in the space-like region~\cite{cello}, while the most accurate measurements from $\pi^0$ decays have an uncertainty of 0.03. Due to the limited precision on the radiative corrections to the $\pi^0_D$ decay~\cite{ka06}, the background description cannot rely on inputs from either experiment or theory.\footnote{Theoretical understanding of radiative corrections to the $\pi^0_D$ decay has been advanced recently, including the final-state radiation~\cite{hu15}. However this development took place after the completion of the present analysis.} An effective TFF slope is therefore obtained from a fit to the measured $m_{ee}$ spectrum itself to provide a satisfactory background description (as quantified by a $\chi^2$ test) in the whole kinematic range $m_{ee}>8~{\rm MeV}/c^2$. The low $m_{ee}$ region is not considered for the DP search due to the significant uncertainties on the acceptance.


\section{Event reconstruction and selection}
\label{sec:selection}


Event selections for the $K_{2\pi}$ and $K_{\mu3}$ decays followed by the prompt $\pi^0\to\gamma A'$, $A'\to e^+e^-$ decay chain are employed. These two selections are identical up to the momentum, invariant mass and particle identification conditions. The principal selection criteria are listed below.
\begin{itemize}
\item Three-track vertices are reconstructed by extrapolation of track segments from the upstream part of the spectrometer into the decay volume, taking into account the measured Earth's magnetic field, stray fields due to magnetization of the vacuum tank, and multiple scattering.
\item The presence of a three-track vertex formed by a pion ($\pi^\pm$) or muon ($\mu^\pm$) candidate and two opposite sign electron ($e^\pm$) candidates is required. Particle identification is based on energy deposition in the LKr calorimeter ($E$) and momentum measured by the spectrometer ($p$). Pions from $K_{2\pi}$ decays and muons from $K_{\mu3}$ decays are kinematically constrained to the momentum range above 5~GeV/$c$, while the momentum spectra of electrons originating from $\pi^0$ decays are soft, peaking at 3~GeV/$c$. Therefore, $p>5~{\rm GeV}/c$ and $E/p<0.85$ ($E/p<0.4$) are required for the pion (muon) candidate, while $p>2.75~{\rm GeV}/c$ and $(E/p)_{\rm min}<E/p<1.15$, where $(E/p)_{\rm min}=0.80$ for $p<5~{\rm GeV}/c$ and $(E/p)_{\rm min}=0.85$ otherwise, are required for the electron candidates. The lower momentum cut and the weaker $E/p$ cut for low momentum electrons are optimised to compensate for the degraded energy resolution (as quantified in Section~\ref{sec:experiment}). The electron identification inefficiency decreases with momentum and does not exceed 0.5\% in the signal momentum range, while the muon identification inefficiency is below 0.1\%. The pion identification inefficiency varies between 1\% and 2\% depending on momentum, and is applied to the simulation using measurements from data samples of fully reconstructed $K_{2\pi}$ and $K^\pm\to 3\pi^\pm$ decays.
\item The tracks forming the vertex are required to be in the fiducial geometric acceptances of the DCH, HOD and LKr detectors. Track separations in the DCH1 plane should exceed 2~cm to reject photon conversions, and electron track separations from electron (pion, muon) tracks in the LKr front plane should exceed 10~cm (25~cm) to minimize the effects of shower overlap.
\item A single isolated LKr energy deposition cluster is considered as a photon candidate. It should be compatible in time with the tracks, and separated by at least 10~cm (25~cm) from the electron (pion, muon) impact points. The reconstructed photon energy should be above 3~GeV to reduce the effects of non-linearity (which is about 1\% at 3~GeV energy) and degraded resolution at low energy.
\item An event is classified as a $K_{2\pi}$ or $K_{\mu3}$ candidate based on the presence of a pion or a muon candidate and the following criteria. The total reconstructed momentum of the three tracks and the photon candidate should be in the range from 53 to 67~GeV/$c$ (below 62~GeV/$c$) for the $K_{2\pi}$ ($K_{\mu 3}$) candidates. The squared total reconstructed transverse momentum with respect to the nominal beam axis ($p_T^2$) should be below $5\times 10^{-4}$~$({\rm GeV}/c)^2$ for the $K_{2\pi}$ candidates, and in the range from $5\times 10^{-4}$ to 0.04~$({\rm GeV}/c)^2$ for the $K_{\mu 3}$ candidates. The two $p_T^2$ intervals do not overlap, therefore the $K_{2\pi}$ and $K_{\mu3}$ event selections are mutually exclusive.
\item The reconstructed invariant mass of the $e^+e^-\gamma$ system is required to be compatible with the nominal $\pi^0$ mass $m_{\pi^0}$~\cite{pdg}: $|m_{ee\gamma}-m_{\pi^0}|<8~{\rm MeV}/c^2$. This interval corresponds to $\pm 5$ times the resolution on $m_{ee\gamma}$.
\item For the $K_{2\pi}$ selection, the reconstructed invariant mass of the $\pi^\pm e^+e^-\gamma$ system should be compatible with the nominal $K^\pm$ mass~\cite{pdg}: $474~{\rm MeV}/c^2<m_{\pi ee\gamma}<514~{\rm MeV}/c^2$. For the $K_{\mu 3}$ selection, the squared missing mass $m_{\rm miss}^2=(P_K-P_\mu-P_{\pi^0})^2$, where $P_\mu$ and $P_{\pi^0}$ are the reconstructed $\mu^\pm$ and $\pi^0$ four-momenta, and $P_K$ is the nominal kaon four-momentum, should be compatible to the missing neutrino mass: $|m_{\rm miss}^2|<0.01~{\rm GeV}^2/c^4$. The resolutions on $m_{\pi ee\gamma}$ and $m_{\rm miss}^2$ are 4.0~MeV/$c^2$ and $1.6\times 10^{-3}~{\rm GeV}^2/c^4$, respectively.
\item The DP mass cut: $|m_{ee}-m_{A'}|<\Delta m(m_{A'})$, where $m_{A'}$ is the assumed DP mass, and $\Delta m(m_{A'})$ is the half-width of the DP search window depending on $m_{A'}$ defined in Section~\ref{sec:dp}.
\end{itemize}

In addition to the above {\it individual DP selections} for the $K_{2\pi}$ and $K_{\mu3}$ decays, the {\it joint DP selection} is also considered: an event passes the joint selection if it passes either the $K_{2\pi}$ or the $K_{\mu3}$ selection. The acceptance of the joint selection $A_{\rm DP}$ for any process is equal to the sum of acceptances of the two mutually exclusive individual selections. Additionally, the {\it Dalitz decay} selections for the $K_{2\pi D}$ and $K_{\mu3D}$ decays are considered: they differ from the DP selections by the absence of the DP mass cut.


\begin{table}[p]
\begin{center}
\caption{Numbers of data events passing the $K_{2\pi D}$ and $K_{\mu 3D}$ selections, and  acceptances of these selections evaluated with MC simulations. The statistical errors on the acceptances are negligible.\label{tab:sel}}
\vspace{2mm}
\begin{tabular}{lcc}
\hline
& $K_{2\pi D}$ selection& $K_{\mu 3D}$ selection \\ 
\hline
\rule{0pt}{11pt}%
Data candidates: & $N_{2\pi D} = 1.38\times 10^7$ & $N_{\mu 3D} = 0.31\times 10^7$ \\
\hline
Acceptances:\\
for $K_{2\pi D}$ decay & $A_\pi(K_{2\pi D})=3.71\%$ & $A_\mu(K_{2\pi D})=0.11\%$ \\ 
for $K_{\mu 3D}$ decay & $A_\pi(K_{\mu3 D})=0.03\%$ & $A_\mu(K_{\mu3 D})=4.17\%$ \\ 
for $K_{3\pi D}$ decay & $A_\pi(K_{3\pi D})=0\phantom{.03\%}$ & $A_\mu(K_{3\pi D})=0.06\%$ \\ 
\hline
\end{tabular}
\end{center}
\end{table}

\begin{figure}[p]
\begin{center}
\resizebox{0.5\textwidth}{!}{\includegraphics{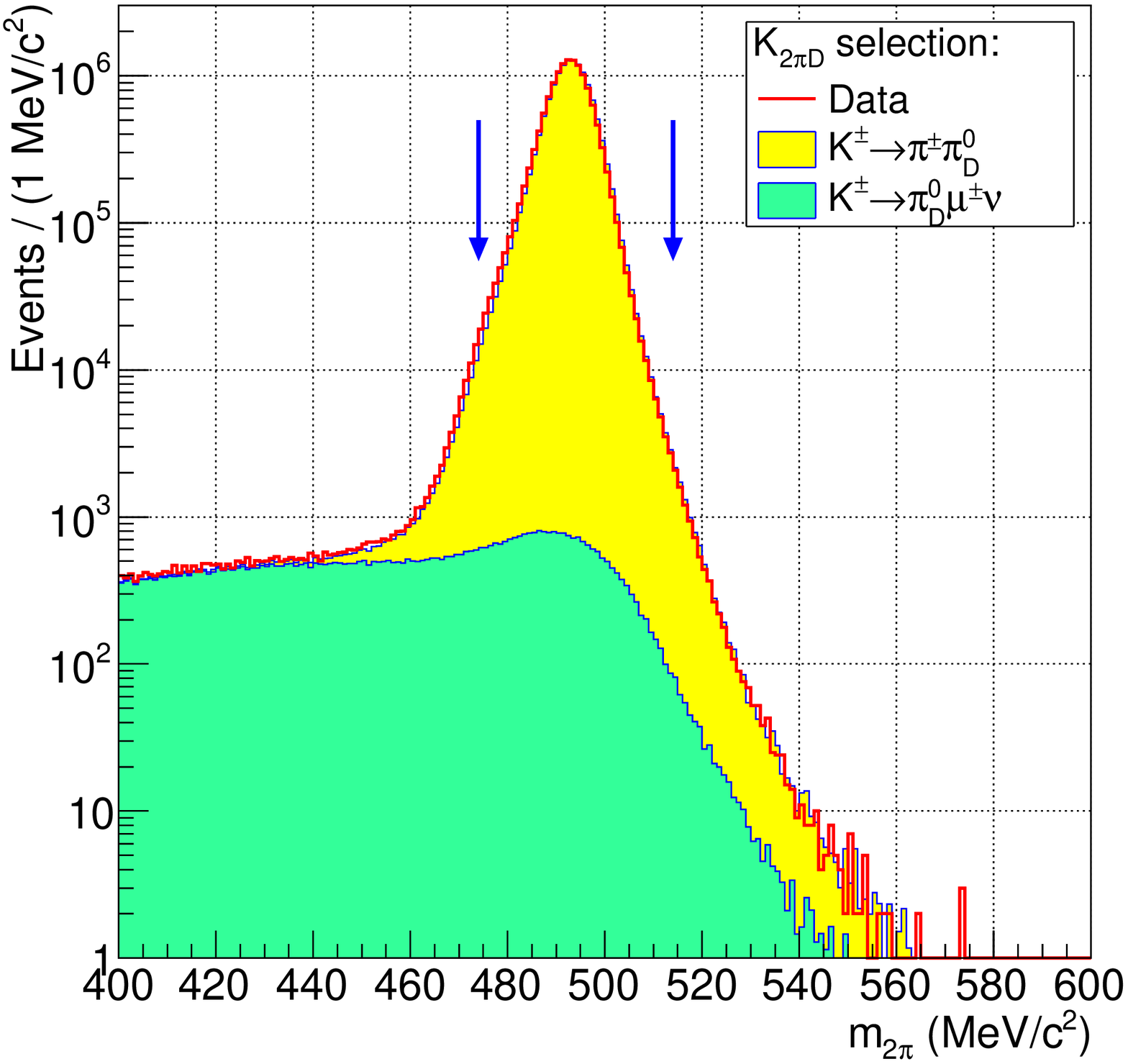}}%
\resizebox{0.5\textwidth}{!}{\includegraphics{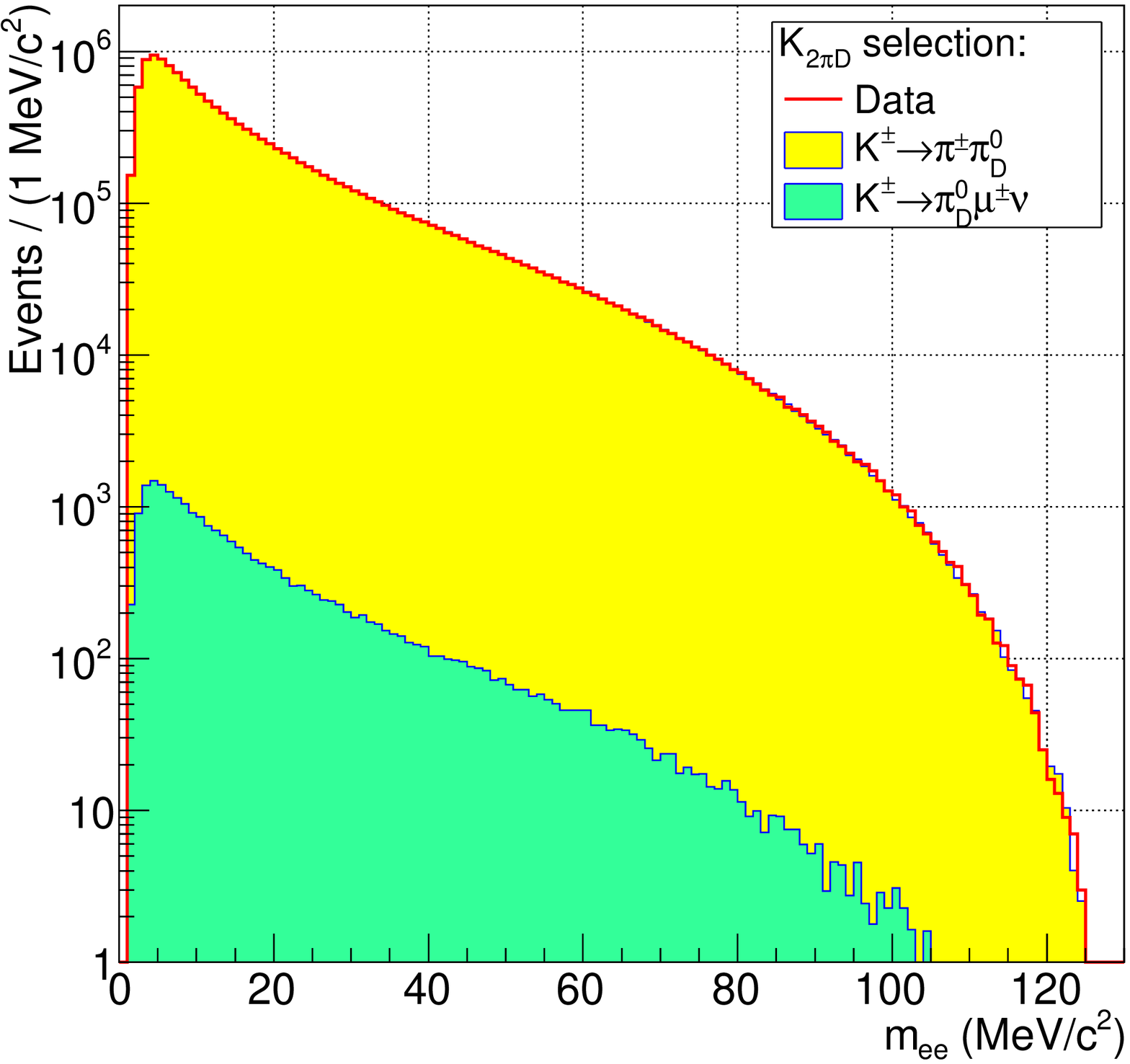}}\\
\resizebox{0.5\textwidth}{!}{\includegraphics{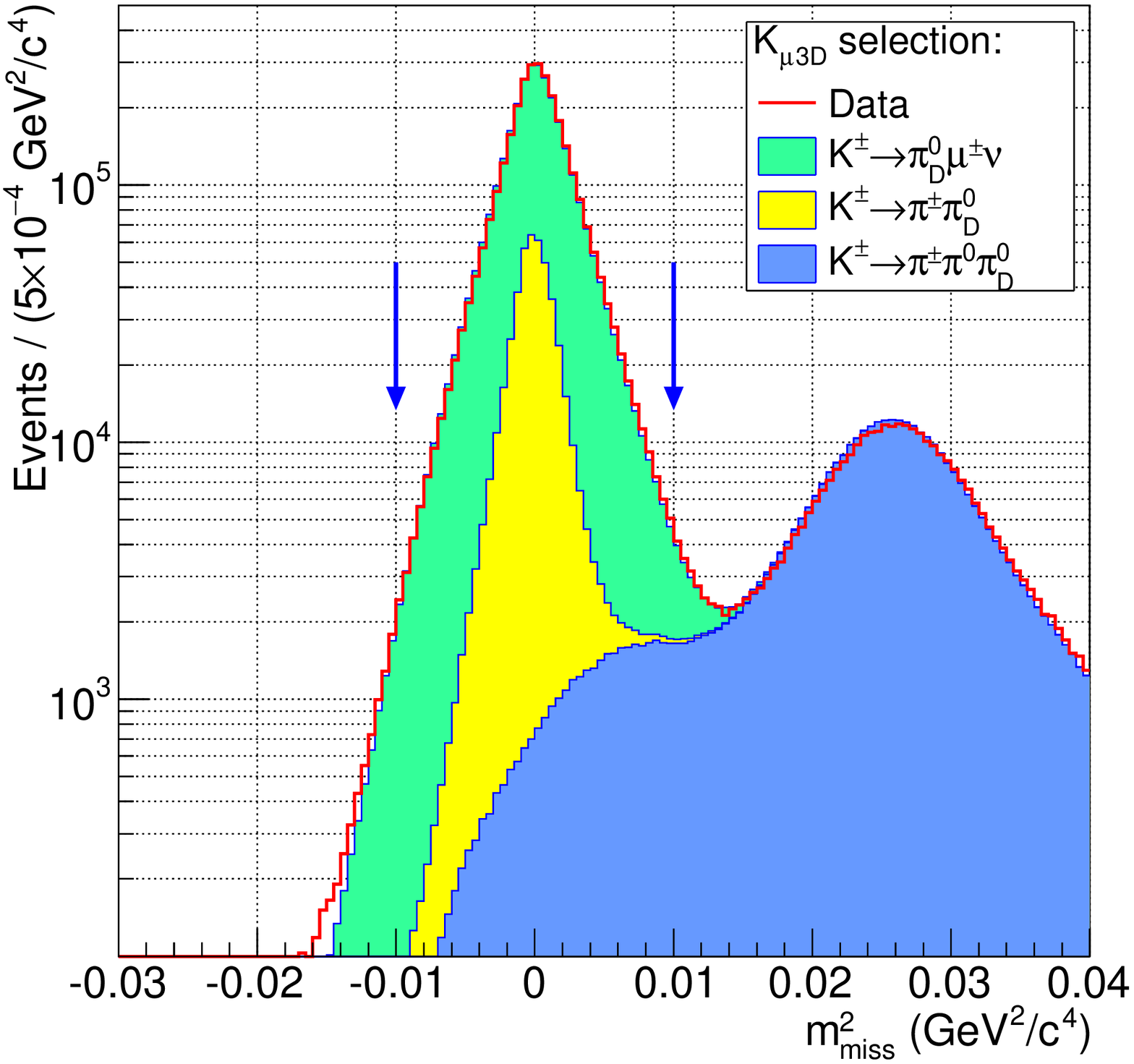}}%
\resizebox{0.5\textwidth}{!}{\includegraphics{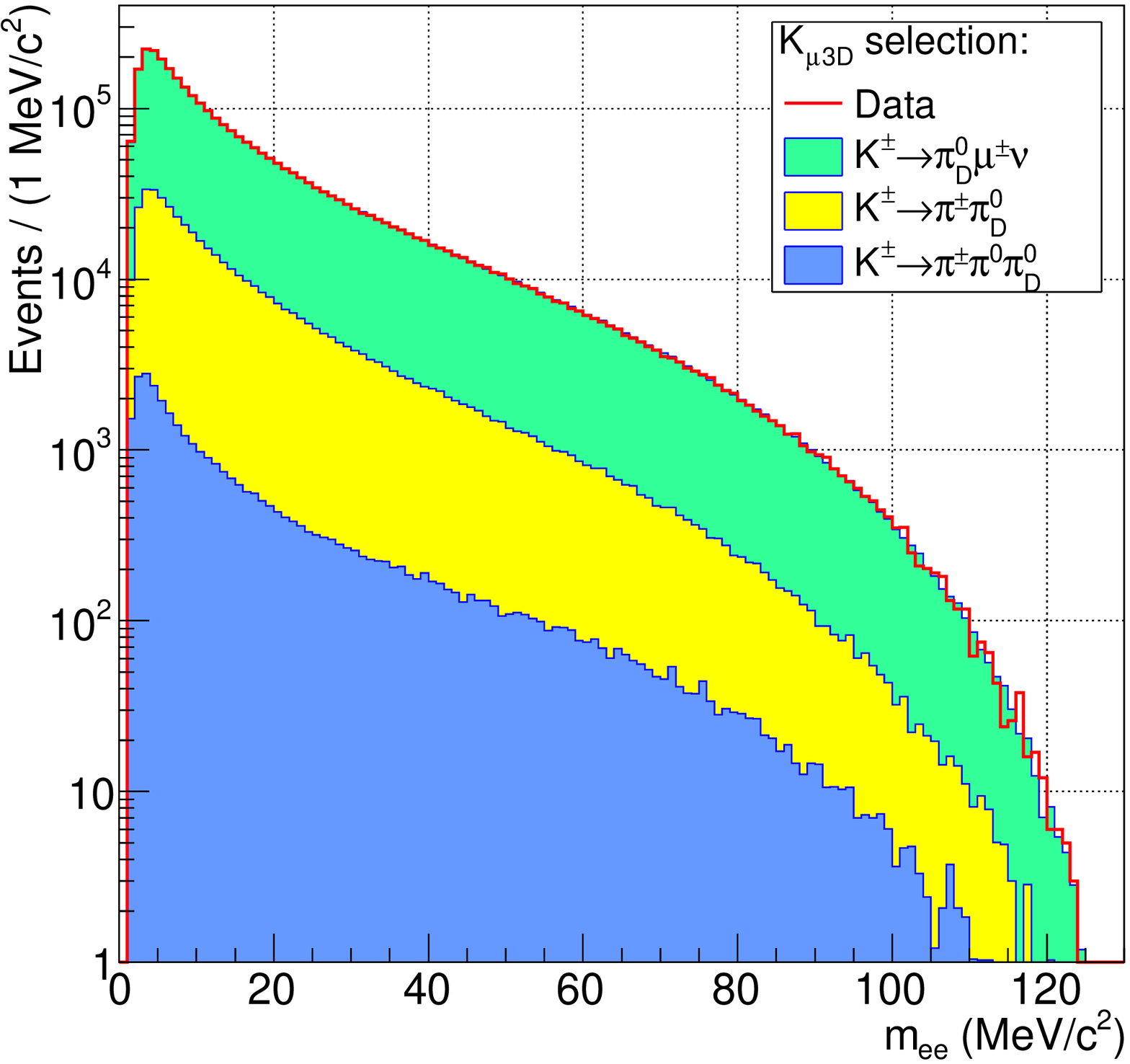}}
\end{center}
\vspace{-7mm}
\caption{Invariant mass distributions of data and MC events passing the $K_{2\pi D}$ (top row) and
$K_{\mu3 D}$ (bottom row) selections. The signal mass regions are indicated with vertical arrows. A dark photon signal would correspond to a spike in the $m_{ee}$ distributions (right column).}
\label{fig:mass}
\end{figure}

\boldmath
\section{Integrated kaon flux and $\pi^0_D$ data sample}
\unboldmath
\label{sec:flux}

The number of $K^\pm$ decays in the 98~m long fiducial decay region is computed as
\begin{equation}
N_K = \frac{N_{2\pi D}}{\left[{\cal B}(K_{2\pi})A_\pi(K_{2\pi D}) + {\cal B}(K_{\mu3})A_\pi(K_{\mu3D})\right] {\cal B}(\pi^0_D) e_1 e_2} =
(1.57\pm0.05)\times 10^{11},
\end{equation}
where $N_{2\pi D}$ is the number of data candidates reconstructed within the $K_{2\pi D}$ selection, $A_\pi(K_{2\pi D})$ and $A_\pi(K_{\mu 3D})$ are the acceptances of the $K_{2\pi D}$ selection for the $K_{2\pi D}$ and $K_{\mu 3D}$ decays evaluated with MC simulations, ${\cal B}(K_{2\pi})$, ${\cal B}(K_{\mu3})$, ${\cal B}(\pi^0_D)$ are the nominal branching fractions of the involved decay modes~\cite{pdg}, and $e_1=(99.75\pm 0.01)\%$, $e_2=(97.50\pm 0.04)\%$ are the efficiencies of the L1 and L2 trigger algorithms measured from downscaled control samples collected simultaneously with the main data set. All numerical quantities are summarized in Table~\ref{tab:sel}. The number of $\pi^0_D$ candidates reconstructed with the joint Dalitz decay selection is $1.69\times 10^7$. The uncertainty on $N_K$ is dominated by the limited precision on ${\cal B}(\pi^0_D)$.

The analysis takes into account the cross-feeding between decay modes. In particular, 0.7\% of the reconstructed $K_{\mu 3D}$ events are classified as $K_{2\pi D}$ due to the low neutrino momentum. Conversely, about 3\% of the reconstructed $K_{2\pi D}$ events are classified as $K_{\mu 3D}$ due to $\pi^\pm\to\mu^\pm\nu$ decays in flight. $K_{3\pi D}$ decays constitute about 1\% of the $K_{\mu 3D}$ candidates.

The reconstructed invariant mass spectra ($m_{\pi ee\gamma}$, $m_{\rm miss}^2$ and $m_{ee}$) of data and MC events passing the Dalitz decay selections, with the MC samples normalised to the data using the estimated value of $N_K$, are shown in Fig.~\ref{fig:mass}.


\section{Search for the dark photon signal}
\label{sec:dp}

A scan for a DP signal in the mass range $9~{\rm MeV}/c^2 \le m_{A'} < 120~{\rm MeV}/c^2$ is performed. The lower boundary of the mass range is determined by the limited accuracy of the $\pi^0_D$ background simulation at low $e^+e^-$ mass. At high DP mass approaching the upper limit of the mass range, the sensitivity to the mixing parameter $\varepsilon^2$ is not competitive with the existing limits due to the kinematic suppression of the $\pi^0\to\gamma A'$ decay.

The resolution on $m_{ee}$ as a function of $m_{ee}$ evaluated with MC simulation is parameterized as $\sigma_{m}(m_{ee}) = 0.067~{\rm MeV}/c^2 + 0.0105\cdot m_{ee}$, and varies from $0.16~{\rm MeV}/c^2$ to $1.33~{\rm MeV}/c^2$ over the mass range of the scan. The intrinsic DP width $\Gamma_{A'}$ is negligible with respect to $\sigma_{m}$. The mass step of the scan and the half-width of the DP search window are defined, depending on the value of $A'$ mass, as $\sigma_{m}(m_{A'})/2$ and $\Delta m=1.5\sigma_m(m_{A'})$, respectively (and both are rounded to the nearest multiple of 0.02 MeV/$c^2$). The search window width has been optimised with MC simulations to achieve the highest expected sensitivity to the DP signal, determined by a trade-off between $\pi^0_D$ background fluctuation and signal acceptance. In total, 404 DP mass values are tested.

\begin{figure}[t]
\begin{center}
\resizebox{0.5\textwidth}{!}{\includegraphics{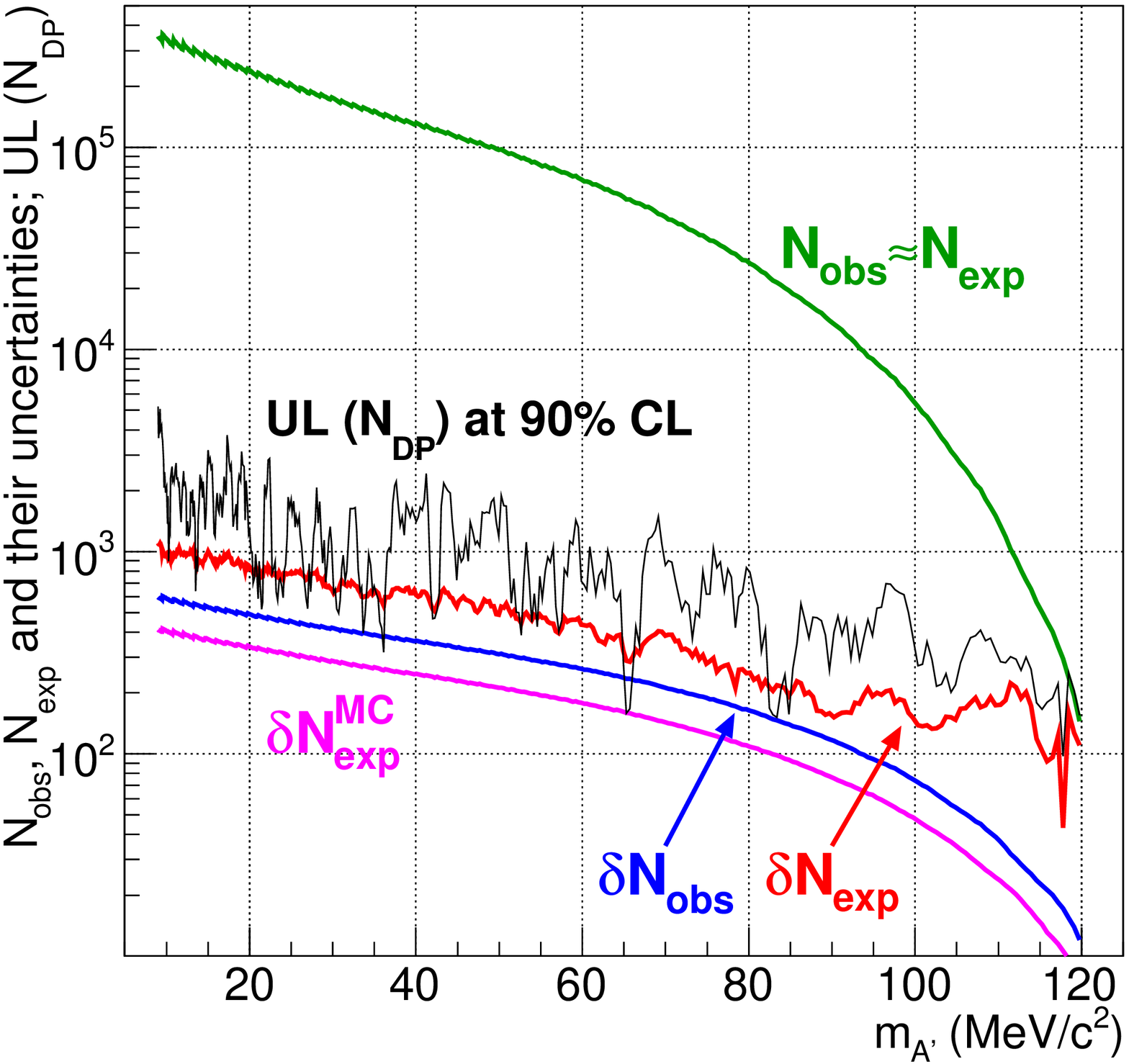}}%
\resizebox{0.5\textwidth}{!}{\includegraphics{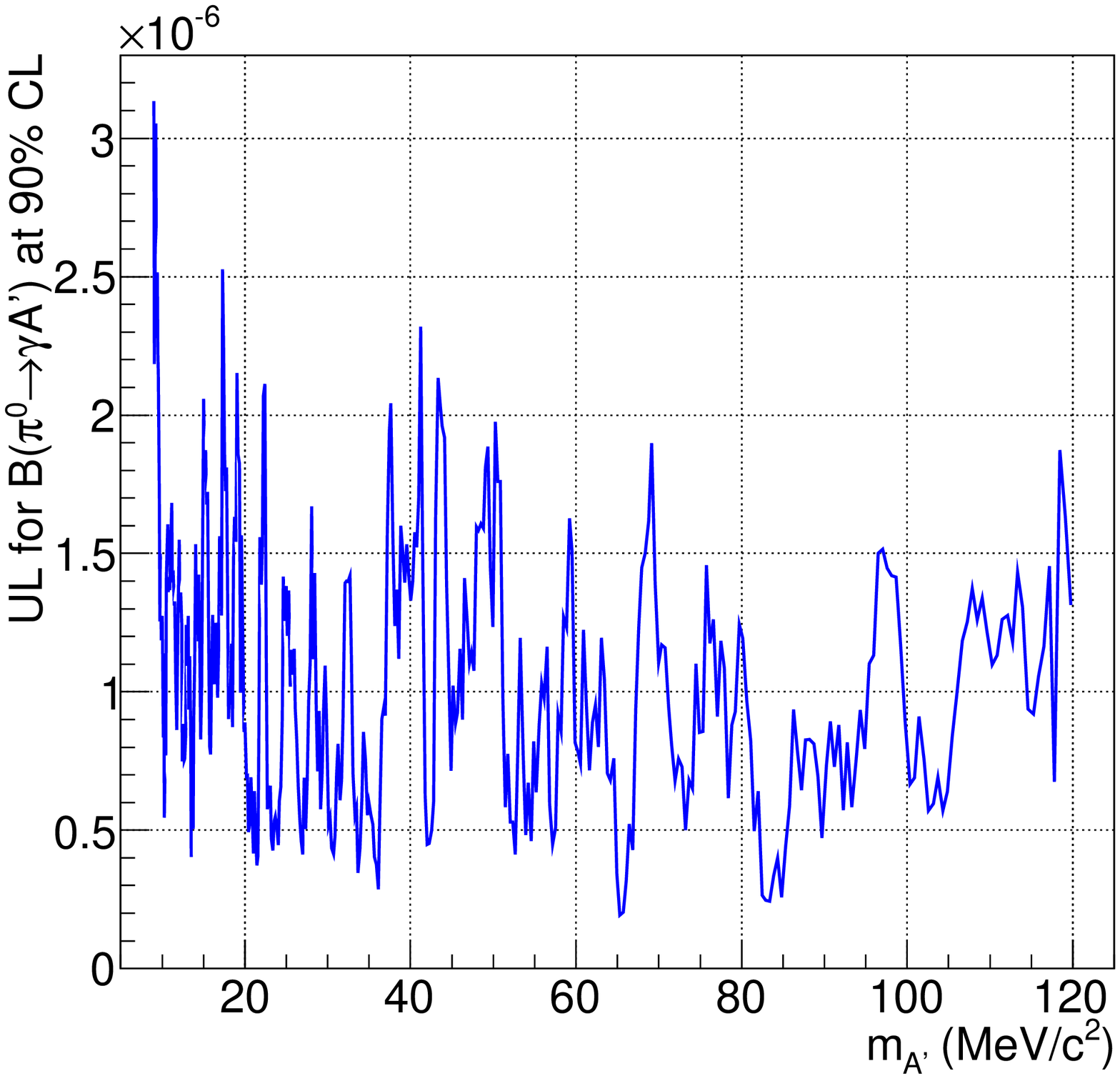}}%
\end{center}
\vspace{-7mm}
\caption{Left: numbers of observed data events ($N_{\rm obs}$) and expected $\pi^0_D$ background events ($N_{\rm exp}$) passing the joint DP selection (indistinguishable in a logarithmic scale), estimated uncertainties $\delta N_{\rm obs}=\sqrt{N_{\rm exp}}$ and $\delta N_{\rm exp}$, and obtained upper limits at 90\% CL on the numbers of DP candidates ($N_{\rm DP}$) for each DP mass value $m_{A'}$. The contribution to $\delta N_{\rm exp}$ from the MC statistical uncertainty is shown separately ($\delta N_{\exp}^{\rm MC}$). The remaining and dominant component is due to the statistical errors on the trigger efficiencies measured in the DP signal region. Right: obtained upper limits on ${\cal B}(\pi^0\to\gamma A')$ at 90\% CL for each DP mass value $m_{A'}$.}
\label{fig:observed}
\end{figure}

For each considered DP mass value, the number of observed data events $N_{\rm obs}$ passing the joint DP selection is compared to the expected number of background events $N_{\rm exp}$. The latter is evaluated from MC simulations, taking into account the trigger efficiency measured from control data samples passing the joint DP selection. The numbers of observed and expected events for each DP mass value and their estimated uncertainties $\delta N_{\rm obs}$ and $\delta N_{\rm exp}$ are shown in Fig.~\ref{fig:observed}. The uncertainty $\delta N_{\rm obs}=\sqrt{N_{\rm exp}}$ is statistical, while the uncertainty $\delta N_{\rm exp}$ has contributions from the limited size of the generated MC samples and the statistical errors on the trigger efficiencies measured in the DP signal region.

The local statistical significance of the DP signal for each mass value estimated as
\begin{equation}
\label{eq:significance}
Z=(N_{\rm obs}-N_{\rm exp})/\sqrt{(\delta N_{\rm obs})^2+(\delta N_{\rm exp})^2}
\end{equation}
never exceeds $3\sigma$, therefore no DP signal is observed. Confidence intervals at 90\% CL for the number of $A'\to e^+e^-$ decay candidates for each DP mass value ($N_{\rm DP}$) are computed from $N_{\rm obs}$, $N_{\rm exp}$ and $\delta N_{\rm exp}$ using the frequentist Rolke--L\'opez method~\cite{ro01}. The obtained upper limits on $N_{\rm DP}$ at 90\% CL are displayed in Fig.~\ref{fig:observed}.

Upper limits at 90\% CL on the branching fraction ${\cal B}(\pi^0\to\gamma A')$ for each DP mass value with the assumption ${\cal B}(A'\to e^+e^-)=1$ (which is a good approximation for $m_A'<2m_\mu$ if $A'$ decays to SM fermions only) are computed using the relation
\begin{equation}
\label{eq:brdp}
{\cal B}(\pi^0\to\gamma A') = \frac{N_{\rm DP}}{N_K e_1 e_2
[{\cal B}(K_{2\pi}) A_{\rm DP}(K_{2\pi}) + {\cal B}(K_{\mu 3}) A_{\rm DP}(K_{\mu 3}) +
2 {\cal B}(K_{3\pi})A_{\rm DP}(K_{3\pi})]},
\end{equation}
where $A_{\rm DP}(K_{2\pi})$, $A_{\rm DP}(K_{\mu 3})$ and $A_{\rm DP}(K_{3\pi})$ are the acceptances of the joint DP selection for $K_{2\pi}$, $K_{\mu 3}$ and $K_{3\pi}$ decays, respectively, followed by the prompt $\pi^0\to\gamma A'$, $A'\to e^+e^-$ decay chain. The trigger efficiencies $e_1$ and $e_2$ (Section~\ref{sec:flux}) are taken into account neglecting their variations over the $m_{ee}$ mass, variations measured to be at the level of a few permille.

Distributions of the angle between the $e^+$ momentum in the $e^+e^-$ rest frame and the $e^+e^-$ momentum in the $\pi^0$ rest frame are identical for the decay chain involving the DP ($\pi^0\to\gamma A'$, $A'\to e^+e^-$) and the $\pi^0_D$ decay, up to the radiative corrections relevant in the latter case but not in the former case. Therefore the acceptances for each DP mass value are evaluated with MC samples of $\pi^0_D$ decays simulated without radiative corrections. The second (third) term in the denominator of Eq.~\ref{eq:brdp} is typically about 20\% (less than 1\%) of the first term. The resulting upper limits on ${\cal B}(\pi^0\to\gamma A')$ are shown in Fig.~\ref{fig:observed}. They are ${\cal O}(10^{-6})$ and do not exhibit a strong dependence on the DP mass, as the mass dependences of $\pi^0_D$ background level and signal acceptances largely compensate each other.

\begin{figure}[t]
\begin{center}
\resizebox{0.55\textwidth}{!}{\includegraphics{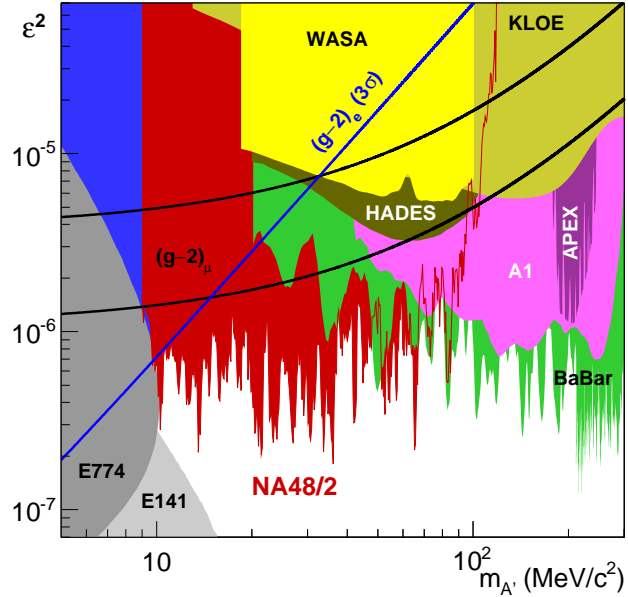}}
\end{center}
\vspace{-7mm}
\caption{Obtained upper limits at 90\% CL on the mixing parameter $\varepsilon^2$ versus the DP mass
$m_{A'}$, compared to other published exclusion limits from meson decay, beam dump and $e^+e^-$ collider experiments. Also shown is the band where the inconsistency of theoretical and experimental values of muon $(g-2)$ reduces to less than 2 standard deviations, as well as the region excluded by the electron $(g-2)$ measurement.}
\label{fig:world}
\end{figure}

Upper limits at 90\% CL on the mixing parameter $\varepsilon^2$ for each DP mass value calculated from the ${\cal B}(\pi^0\to\gamma A')$ upper limits using Eq.~\ref{eq:br} are shown in Fig.~\ref{fig:world}, together with the constraints from other experiments (see Ref.~\cite{ba15} for details). Also shown is the band in the ($m_{A'}$, $\varepsilon^2$) plane where the discrepancy between the measured and calculated muon $(g-2)$ values falls into the $\pm2\sigma$ range due to the DP contribution, as well as the region excluded by the electron $(g-2)$ measurement~\cite{po09,en12,da14}.

The most stringent limits on $\varepsilon^2$ obtained are those at DP mass where the kinematic suppression of the $\pi^0\to\gamma A'$ decay is weak. The prompt DP decay assumption is justified a posteriori by the achieved limits. Given the 60~GeV/$c$ beam, the maximum DP mean path in the laboratory reference frame corresponds to an energy of approximately $E_{\rm max}=50~{\rm GeV}$:
\begin{equation}
L_{\rm max} \approx \frac{E_{\rm max}}{m_{A'}c^2} \cdot c\tau_{A'} \approx 0.4~{\rm mm} \times \left(\frac{10^{-6}}{\varepsilon^2}\right) \times \left(\frac{100~{\rm MeV}/c^2}{m_{A'}}\right)^2.
\end{equation}
The lowest obtained limit $\varepsilon^2 m_{A'}^2=3\times 10^{-5}~{\rm MeV}^2/c^4$ translates into a maximum DP mean path of $L_{\rm max}\approx 10~{\rm cm}$. The corresponding loss of the 3-track trigger and event reconstruction efficiency is negligible, as the offline resolution on the longitudinal coordinate of a 3-track vertex is about 1~m.

The sensitivity of the prompt $A'$ decay search is limited by the irreducible $\pi^0_D$ background. In particular, the upper limits on ${\cal B}(\pi^0\to\gamma A')$ and $\varepsilon^2$ obtained in this analysis are two to three orders of magnitude above the single event sensitivity. The achievable upper limit on $\varepsilon^2$ scales as the inverse square root of the integrated beam flux, which means that the possible improvements to be made with this technique using larger future $K^\pm$ samples (e.g. at the NA62 experiment) are modest. On the other hand, the sensitivity of the displaced $A'$ decay vertex analysis to lower $\varepsilon^2$ is being investigated.


\boldmath
\section{Dark photon search in the $K^\pm\to\pi^\pm A'$ decay}
\unboldmath

\begin{figure}[t]
\begin{center}
\resizebox{0.5\textwidth}{!}{\includegraphics{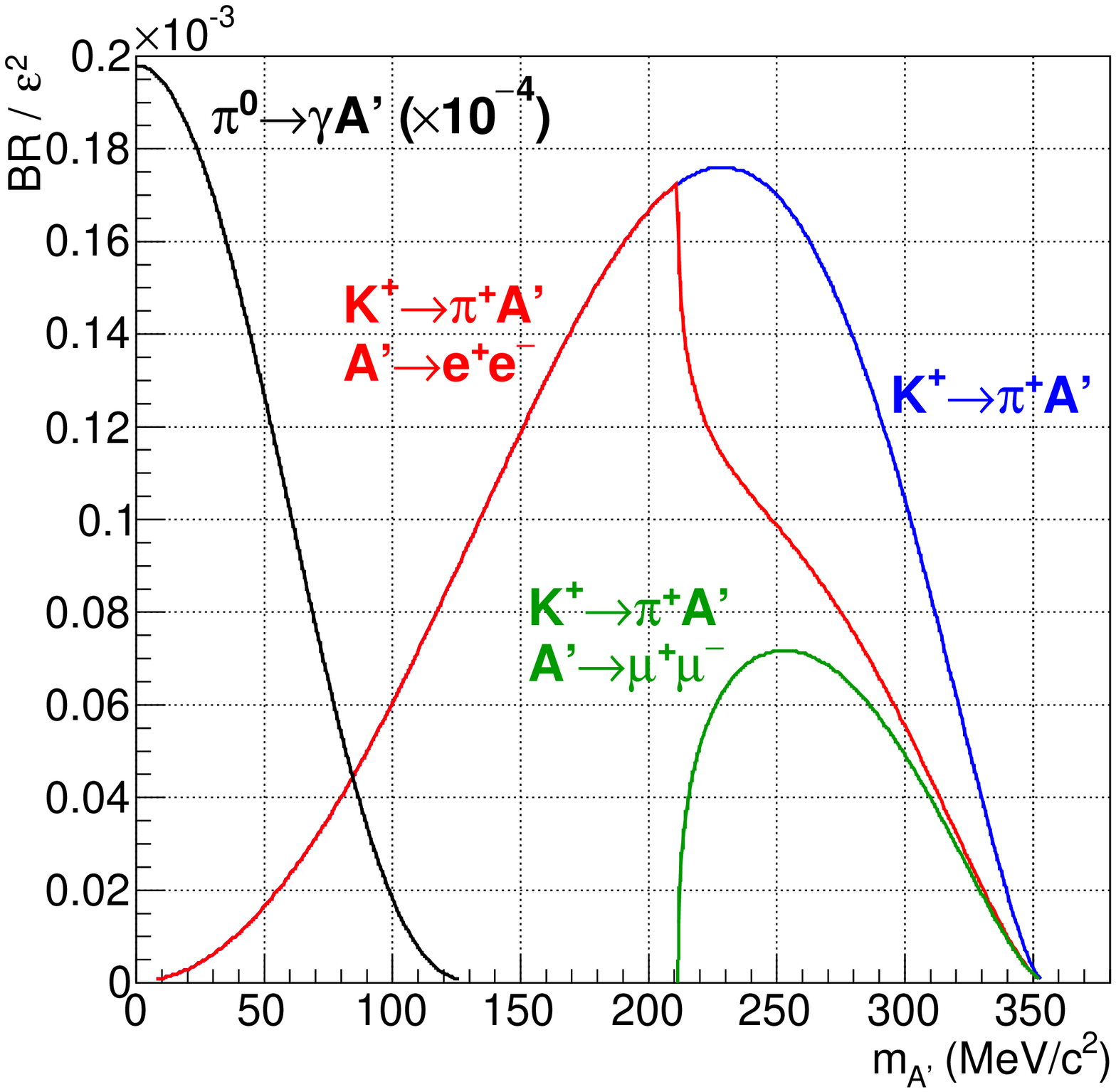}}%
\resizebox{0.5\textwidth}{!}{\includegraphics{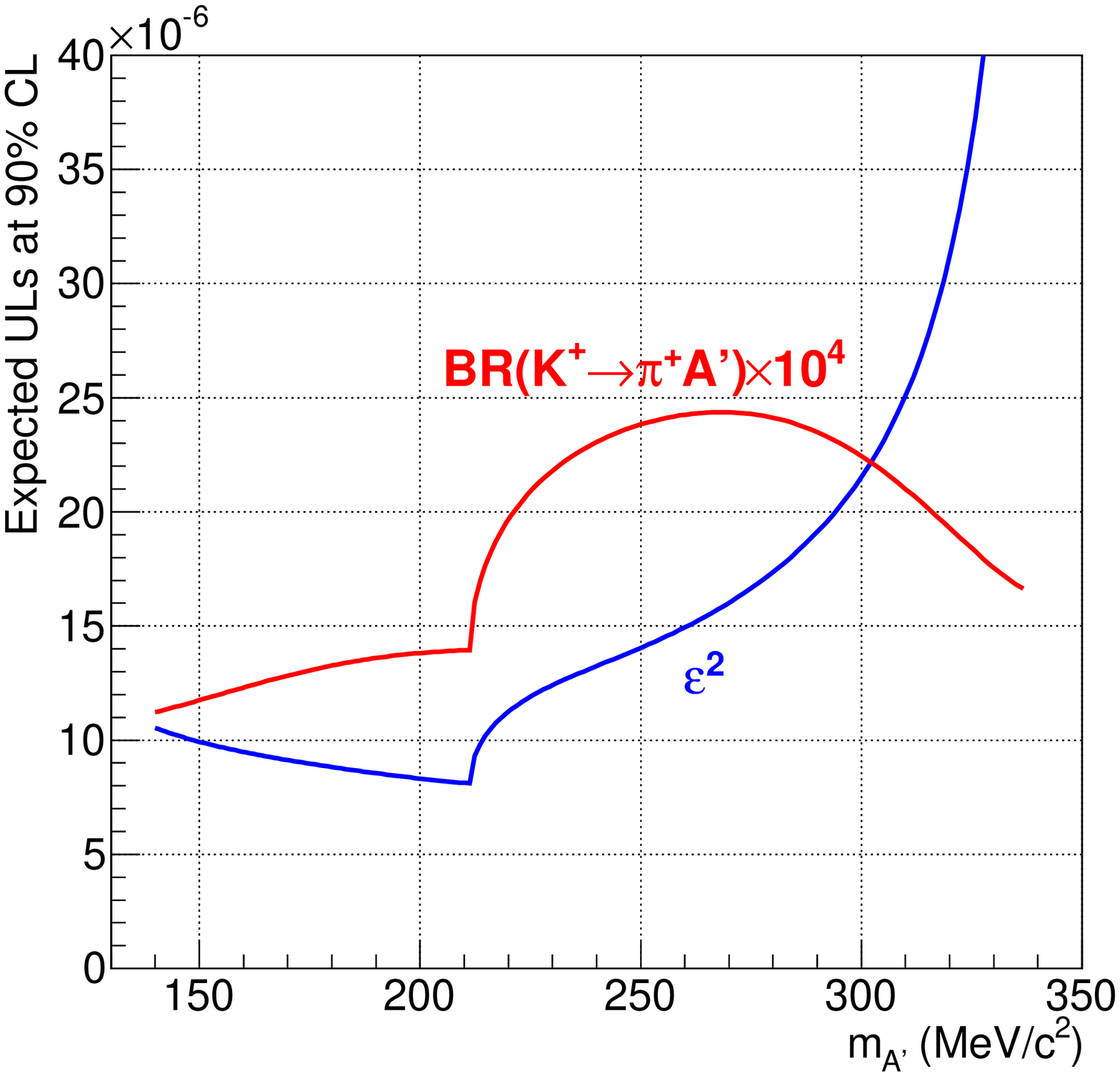}}
\end{center}
\vspace{-7mm}
\caption{Left: branching ratios of $\pi^0$ and $K^+$ decays into the DP normalized to the mixing parameter $\varepsilon^2$~\cite{batell09,da14}. The branching ratio ${\cal B}(\pi^0\to\gamma A')$ is scaled by a factor of $10^{-4}$. The products ${\cal B}(K^\pm\to\pi^\pm A')\times {\cal B}(A'\to\ell^+\ell^-)$ are also shown, assuming that DP decays into SM fermions only. Right: the expected NA48/2 upper limits at 90\% CL for ${\cal B}(K^\pm\to\pi^\pm A')$ and the mixing parameter $\varepsilon^2$, considering the $A'\to e^+e^-$ decay and assuming that DP decays into SM fermions only. Note the drop in sensitivity corresponding to the drop in ${\cal B}(A'\to e^+e^-)$ at the di-muon threshold.}
\label{fig:Kdecays}
\end{figure}

An alternative way to search for the DP in $K^\pm$ decays is via the $K^\pm\to\pi^\pm A'$ decay followed by the prompt $A'\to\ell^+\ell^-$ decay ($\ell=e,\mu$). The experimental signature is identical to those of the well established rare decays $K^\pm\to\pi^\pm\ell^+\ell^-$~\cite{ba09,ba11}. This decay chain provides sensitivity to the DP in the mass range $2m_e<m_{A'}<m_K-m_\pi$; in practice the region $m_{A'}<m_{\pi^0}$ dominated by the $\pi^0_D$ background is not accessible. As seen in Fig.~\ref{fig:Kdecays}, the expected branching fraction is ${\cal B}(K^\pm\to\pi^\pm A') < 2\cdot 10^{-4}\varepsilon^2$ over the whole allowed $m_{A'}$ range, in contrast to ${\cal B}(\pi^0\to\gamma A') \sim \varepsilon^2$ for $m_{A'}<100~{\rm MeV}/c^2$. At NA48/2, the suppression of the DP production in the $K^+$ decay relative to $\pi^0$ decay is partly compensated by the favourable ration of the numbers of $K^\pm$/$\pi^0$ decays ($\approx 3$), much lower background (mainly from rare decays $K^\pm\to\pi^\pm\ell^+\ell^-$ for $m_{A'}>m_{\pi^0}$) and higher acceptance (a factor of $\approx 4$).

The expected sensitivity of the NA48/2 data sample to ${\cal B}(K^\pm\to\pi^\pm A')$ considering the $A'\to e^+e^-$ decay only and assuming decays only into SM fermions, as well as the corresponding sensitivity to $\varepsilon^2$, have been evaluated: the results are presented in Fig.~\ref{fig:Kdecays}. The sensitivity is maximal in the mass interval $140~{\rm MeV}/c^2<m_{A'}<2m_\mu$, where the $K^\pm\to\pi^\pm A'$ decay is not kinematically suppressed, the $\pi^0_D$ background is absent, and ${\cal B}(A'\to e^+e^-)\approx 1$. In this $m_{A'}$ interval, the expected NA48/2 upper limits lie in the range $\varepsilon^2=(0.8-1.1)\times 10^{-5}$ at 90\% CL, in agreement with earlier generic estimates~\cite{po09,da14}. The sensitivity achievable via the $A'\to\mu^+\mu^-$ decay is lower due to the lower $A'\to\mu^+\mu^-$ decay rate. The expected limits displayed in Fig.~\ref{fig:Kdecays} are not competitive with the existing exclusion limits (Fig.~\ref{fig:world}), therefore the corresponding search has not been performed.

An interesting possibility however is the search for the $K^+\to\pi^+A'$, $A'\to {\rm invisible}$ decays. This has been done by studying the spectrum of $K^+\to\pi^+\nu\bar\nu$ candidates observed by the BNL E787 and E949 experiments~\cite{ar09}. The resulting constraints and non-trivial, as most of the constraints presented in Fig.~\ref{fig:world} do not apply in the case of $A'\to {\rm invisible}$~\cite{da14}. The NA62 experiment at CERN is expected to set even more stringent constraints via the $K^+\to\pi^+\nu\bar\nu$ measurement.

\section*{Conclusions}

A search for the dark photon (DP) production in the $\pi^0\to\gamma A'$ decay followed by the prompt $A'\to e^+e^-$ decay has been performed using the data sample collected by the NA48/2 experiment in 2003--2004. No DP signal is observed, providing new and more stringent upper limits on the mixing parameter $\varepsilon^2$ in the mass range 9--70 MeV/$c^2$. In combination with other experimental searches, this result rules out the DP as an explanation for the muon $(g-2)$ measurement under the assumption that the DP couples to quarks and decays predominantly to SM fermions. The prospects for dark photon search via the $K^\pm\to\pi^\pm A'$ decay are briefly discussed.

\end{document}